\documentclass[aps,pre,twocolumn,superscriptaddress]{revtex4}
\usepackage[dvips]{graphics}
\usepackage{graphicx}
\usepackage{amsfonts}
\usepackage{amssymb}
\usepackage{amsmath}
\usepackage{subfigure}

\begin{document}

\title{Stability of Discrete Solitons in the Presence of Parametric Driving}
\author{H. Susanto}
\affiliation{Department of Mathematics and Statistics, University of Massachusetts,
Amherst MA 01003-4515, USA}

\author{Q.E. Hoq}
\affiliation{Department of Mathematics, Western New England College, Springfield, MA, 01119, USA}

\author{P.G.\ Kevrekidis}
\affiliation{Department of Mathematics and Statistics, University of Massachusetts,
Amherst MA 01003-4515, USA}

\begin{abstract}
In this brief report, we consider parametrically driven bright solitons
in the vicinity of the anti-continuum limit. We illustrate the mechanism
through which these solitons become unstable due to the collision of
the phase mode with the continuous spectrum, or eigenvelues bifurcating
thereof. We show how this mechanism typically leads to {\it complete
destruction} of the bright solitary wave.
\end{abstract}

\maketitle


{\it Introduction.} In the past few years, 
differential-difference dispersive equations where  
the evolution 
variable is continuum but the spatial variables are discrete, 
have been the focus of intense research efforts \cite{reviews}.
The key reason for the increasing interest in this research 
direction can be attributed to the wide range of pertinent 
applications ranging, from e.g.,      
the spatial dynamics of optical beams in coupled waveguide arrays in 
nonlinear optics \cite{reviews1}, to 
the temporal evolution of Bose-Einstein condensates (BECs) 
in deep, optically-induced, lattice potentials 
in soft-condensed matter physics \cite{reviews2}, or even to 
the DNA double strand in biophysics \cite{reviews3} among others. 

One of the key models that has emerged in all of the above settings,
either as describing e.g., the envelope wave of the electric field 
in the optical setting \cite{christo}, or describing the wavefunction
at the nodes of the optical lattice in BECs \cite{tromb}, is the
discrete nonlinear Schr{\"o}dinger (DNLS) equation. This prototypical
lattice model features
a dispersive coupling between nearest-neighbors, and a cubic
onsite nonlinearity. 

The above spatially discrete model bears a number of
interesting similarities and differences, in comparison with its continuum
sibling, the famous (integrable in 1-spatial dimension) nonlinear
Schr{\"o}dinger equation (NLS) \cite{sulem}. One of the key 
differences is the breaking of one of the important invariances
of the NLS model, namely of the translational invariance that is 
responsible for momentum conservation in that setting. On the 
contrary, the discrete model carries an integer-shift invariance.
This has some important implications, among other things, to the nature
of the solutions of the discrete model. In fact, it was realized 
through perturbative calculations \cite{campbell} and subsequently
more rigorously justified \cite{kevkap} 
that the principal (single-humped solitary
wave) solutions of the latter model can only be centered on a lattice
site or between two lattice sites. In the continuum case, the center
of the solution is a free parameter due to the continuum invariance.

On the other hand, one of the important similarities of the discrete
model to the continuum one is the presence of the so-called phase
or gauge invariance (which is associated with the overall freedom
of selecting the solution's phase). The conservation law related to
this invariance is the one of the $L^2$ (respectively $l^2$) norm,
or ``mass'' of the solution. This invariance is the main focal
point of the present work. In particular, we introduce, arguably,
the simplest possible perturbation that breaks the relevant invariance,
in the form of a parametric drive. The relevance of such a
term involving a perturbation proportional to the complex conjugate
of the field has been discussed in a variety of earlier works (see
e.g. \cite{barashenkov} and references therein). A specific physical
setting where this type of perturbation arises can be found by
looking at the envelope equation of a system of parametrically 
driven (undamped) coupled torsion pendula as discussed in \cite{bar1}
(with the difference that the envelope wave expansion should be
performed in a genuinely discrete setting similarly to \cite{kivpey}
rather than near the continuum limit as in \cite{bar1}). 
The aim of this exposition is to examine how the breaking of
this invariance results in an eigenvalue that bifurcates from
the origin of the spectral plane, when linearizing around the
most fundamental, solitary wave solution. We argue (analytically and
support numerically) that this eigenvalue
can lead to an instability of the solitary wave for an isolated value
of the parametric drive even at the so-called anti-continuum limit where
lattice sites are uncoupled. For non-vanishing couplings, the same eigenvalue
leads to a wide interval of parametric instabilities in the two-parameter
space (of parametric drive versus inter-site coupling) that we explore
both analytically and numerically. Within this interval, we also elucidate
the typical numerical behavior of the solitary wave solutions, using
direct numerical simulations of relevant unstable waveforms.

Our presentation will be structured as follows. In the next section, we present
our analytical setup and perturbative results. Then,
we compare our analytical findings with the results of numerical computations.
Finally, we summarize our findings and present our conclusions,
as well as motivate some questions for future study.


{\it Setup and perturbation analysis.} 
The model we consider is the perturbed (i.e., parametrically driven)
discrete non-linear Schr\"odinger equation of the form
\begin{equation}
i\dot{\phi}_n=-C\Delta_2\phi_n-|\phi_n|^2\phi_n+\Lambda \phi_n+\gamma \overline{\phi_n},
\label{schr}
\end{equation}
where $C$ is the coupling constant between two adjacent sites of the lattice, $\Delta_2\phi_n=(\phi_{n+1}-2\phi_n+\phi_{n-1})$ 
is the discrete Laplacian, $\Lambda$ is the propagation constant in optics or the chemical potential in BECs, and $\gamma$
is the strength of the parametric drive. 

We focus our attention on a standing wave profile so that $\phi_n$ is time-independent. In this case, $\phi_n$ satisfies
\begin{equation}
-C\Delta_2\phi_n-\phi_n^3+\Lambda \phi_n+\gamma \phi_n=0.
\label{schr2}
\end{equation}

In the uncoupled (or so-called anti-continuum) limit of $C=0$, 
the solution of \eqref{schr2} is $\phi_n=0,\pm\sqrt{\Lambda+\gamma}$. 
We examine here the most fundamental single-hump solitary wave solutions
which in the anti-continuum limit emanate from a single-site excitation
of the form:
\begin{equation}
u_n^0=0,\,n\neq0,\quad u_0^0=\sqrt{\Lambda+\gamma}.
\label{ds0}
\end{equation}

The continuation of \eqref{ds0} for small coupling $C$ can be calculated 
analytically through a perturbative expansion. By substituting into the 
steady state equation \eqref{schr2} 
$u_n=u_n^0+Cu_n^1 + C^2 u_n^{2} + \dots$, one can calculate that up to 
order $\mathcal{O}(C^2)$
\begin{equation}
u_n=\left\{\begin{array}{ll}
\sqrt{\Lambda+\gamma}+C/\sqrt{\Lambda+\gamma},\,&n=0,\\
C/\sqrt{\Lambda+\gamma},\,&n=-1,1,\\
0,\,&n\,\text{otherwise}.
\end{array}
\right.
\label{ds1}
\end{equation}

To perform linear stability analysis to the discrete solitary waves of
the form of Eq. \eqref{ds1}, we introduce the following linearization
ansatz
\[
\phi_n=u_n+ \delta \epsilon_n.
\]
Substituting into \eqref{schr}  yields the following linearized equation
to $\mathcal{O}(\delta)$
\begin{equation}
i\dot{\epsilon}_n=-C\Delta_2\epsilon_n-2|u_n|^2\epsilon_n-u_n^2\overline{\epsilon_n}+\Lambda\epsilon_n+\gamma\overline{\epsilon_n}.
\label{lin}
\end{equation}

Writing $\epsilon_n(t)=\eta_n+i\xi_n$ and assuming that $u_n$ is real, 
eq.\ \eqref{lin} gives (see, e.g., \cite{joha99})
\begin{eqnarray}
\left(\begin{array}{cc}
\dot{\eta}_n\\
\dot{\xi}_n
\end{array}\right)=
\left( 
\begin{array}{cc} 
0 & {\cal L}_+(C) \\
-{\cal L}_-(C) & 0
\end{array} 
\right)
\left(\begin{array}{cc}
{\eta}_n\\
{\xi}_n
\end{array}\right)=
{\cal H}
\left(\begin{array}{cc}
{\eta}_n\\
{\xi}_n
\end{array}\right),
\label{lin2}
\end{eqnarray}
where the operator ${\cal L}_-(C)$ and ${\cal L}_+(C)$ are defined as $\mathcal{L}_-(C)\equiv -C\Delta_2-(3u_n^2-\Lambda-\gamma)$ and $\mathcal{L}_+(C)\equiv -C\Delta_2-(u_n^2-\Lambda+\gamma)$. The stability of $u_n$ is then determined by the eigenvalues of ${\cal H}$.

Let the eigenvalues of ${\cal H}$ be denoted by $i\omega$, which implies that $u_n$ is stable if Im$(\omega)=0$. Because \eqref{lin2} is linear, we can eliminate one of the 'eigenvectors', for instance
$\xi_n$, from which we obtain the following eigenvalue problem
\begin{equation}
{\cal L}_+(C){\cal L}_-(C)\eta_n=\omega^2\eta_n=\Omega\eta_n.
\label{evp}
\end{equation}

As before, we expand the eigenvector $\eta_n$ and the eigenvalue $\Omega$ as 
\[\eta_n=\eta_n^0+C\eta_n^1+{\cal O}(C^2),\quad \Omega=\Omega^0+C\Omega^1+{\cal O}(C^2).\]

Substituting into Eq.\ \eqref{evp} and identifying coefficients for consecutive powers of the small parameter $C$ 
yields
\begin{eqnarray}
\displaystyle \left[{\cal L}_+(0){\cal L}_-(0)-\Omega^0\right]\eta_n^0&=&0,\label{o0}\\
\displaystyle \left[{\cal L}_+(0){\cal L}_-(0)-\Omega^0\right]\eta_n^1&=&f,\label{o1}
\end{eqnarray}
with
\begin{equation}
\displaystyle f=\left[-(\Delta_2+2u^0_nu^1_n){\cal L}_-(0)-{\cal L}_+(0)(\Delta_2+6u_n^0u_n^1)+\Omega^1\right]\eta_n^0.
\label{f}
\end{equation}

First, let us consider the order ${\cal O}(1)$ equation \eqref{o0}.
One can do a simple analysis to show that there are only two eigenvalues, 
i.e., \ $\Omega^0=\Lambda^2-\gamma^2$ 
and $\Omega^0=4(\Lambda+\gamma)\gamma$. $\Omega^0=\Lambda^2-\gamma^2$ has infinite multiplicity and is related to 
the continuous spectrum that will be discussed later. Therefore, our 
interest is in $\Omega^0=4(\Lambda+\gamma)\gamma$
that has the normalized eigenvector $\eta^0_n=0,\,n\neq0$ and $\eta^0_0=1$. This eigenvalue is the formerly 
zero eigenvalue due to the phase or gauge invariance of the DNLS equation
{\it in the absence of parametric driving}.

The continuation of the eigenvalue $\Omega^0=4(\Lambda+\gamma)\gamma$ when the coupling $C$ is turned on can be calculated
from Eq.\ \eqref{o1}. Due to the corresponding eigenvector having 
$\eta^0_n=0$ for $n\neq0$, we only 
need to 
consider the site $n=0$. In this case, $f=-8\gamma+\Omega^1$. 
The solvability condition of Eq.\ \eqref{o1} using, e.g., 
the Fredholm alternative requires $f=0$ from which 
we immeadiately obtain that $\Omega^1=8\gamma$. 
Hence, the smallest eigenvalue of a one-site discrete soliton solution of Eq.\ 
\eqref{schr} is
\begin{equation}
\Omega = 4(\Lambda+\gamma)\gamma + 8\gamma C+{\cal O}(C^2),
\label{Wg}
\end{equation}
or
\begin{equation}
\omega = \pm2\sqrt{(\Lambda+\gamma)\gamma} \pm 2\frac{\gamma}{\sqrt{(\Lambda+\gamma)\gamma}}C+{\cal O}(C^2).
\label{wg}
\end{equation}

\begin{figure}[tb]
\centerline{
\includegraphics[height=6cm,angle=0,clip]{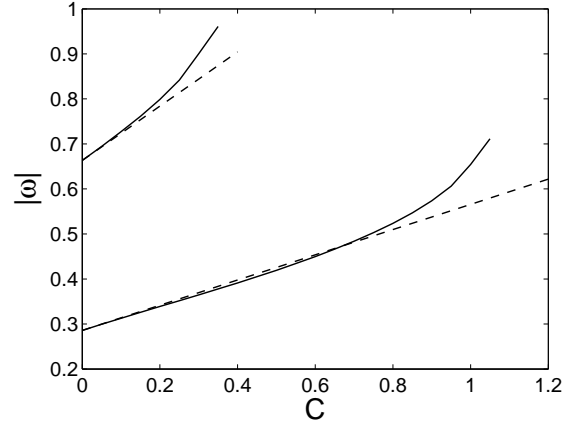}
}
\caption{The smallest eigenvalue for two values of $\gamma$, namely
\ $\gamma=0.02$ and $\gamma=0.1$. The dashed lines are the approximate 
analytical estimate of the relevant frequency from Eq. \eqref{wg}. The
lower curves correspond to $\gamma=0.02$.} 
\label{Fig1}
\end{figure}

\begin{figure}[tb]
\centerline{
\includegraphics[height=6cm,angle=0,clip]{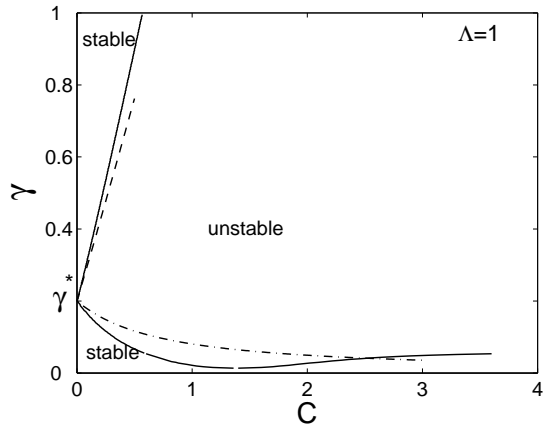}
}
\caption{The stability-instability region in the 
two-parameter space $\gamma-C$. The solid lines give the 
numerically obtained separatrices, while the 
dash-dotted and dashed ones the analytical approximations of Eqs.
(\ref{gc1}) and (\ref{gc2}) respectively.} 
\label{Fig2}
\end{figure}

\begin{figure}[tb]
\subfigure{\includegraphics[height=3cm,angle=0,clip]{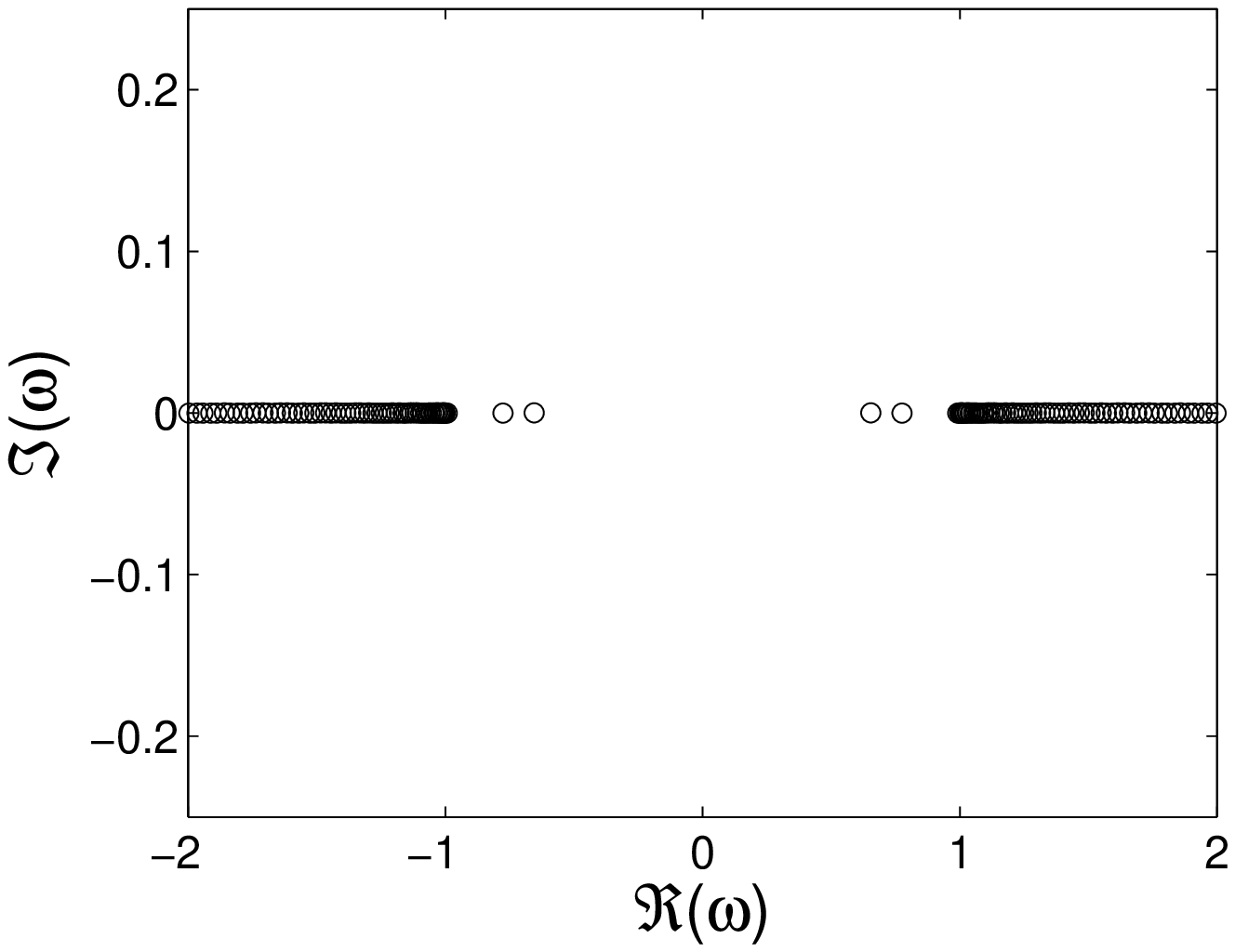}}
\subfigure{\includegraphics[height=3cm,angle=0,clip]{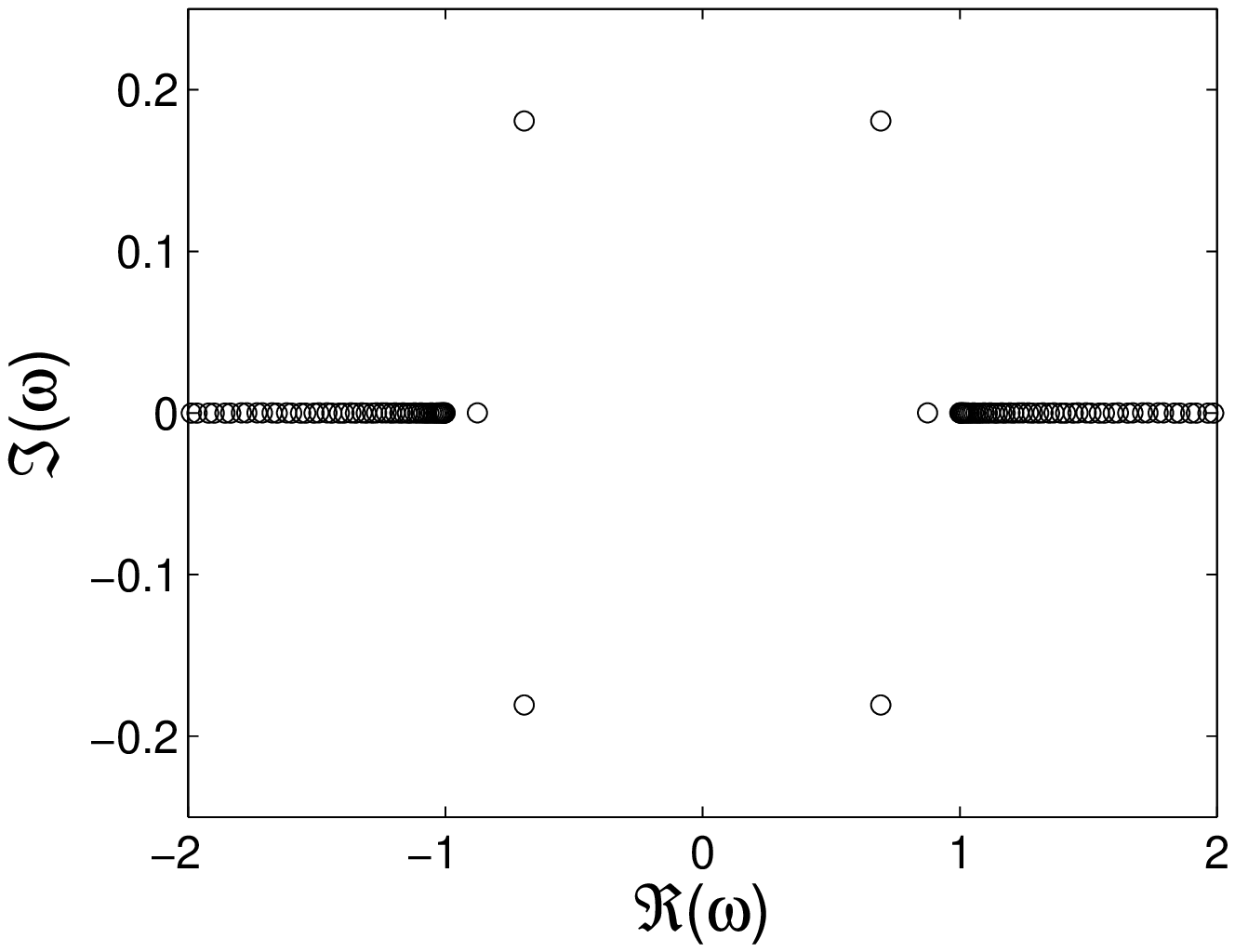}}\\
\subfigure{\includegraphics[height=5cm,angle=0,clip]{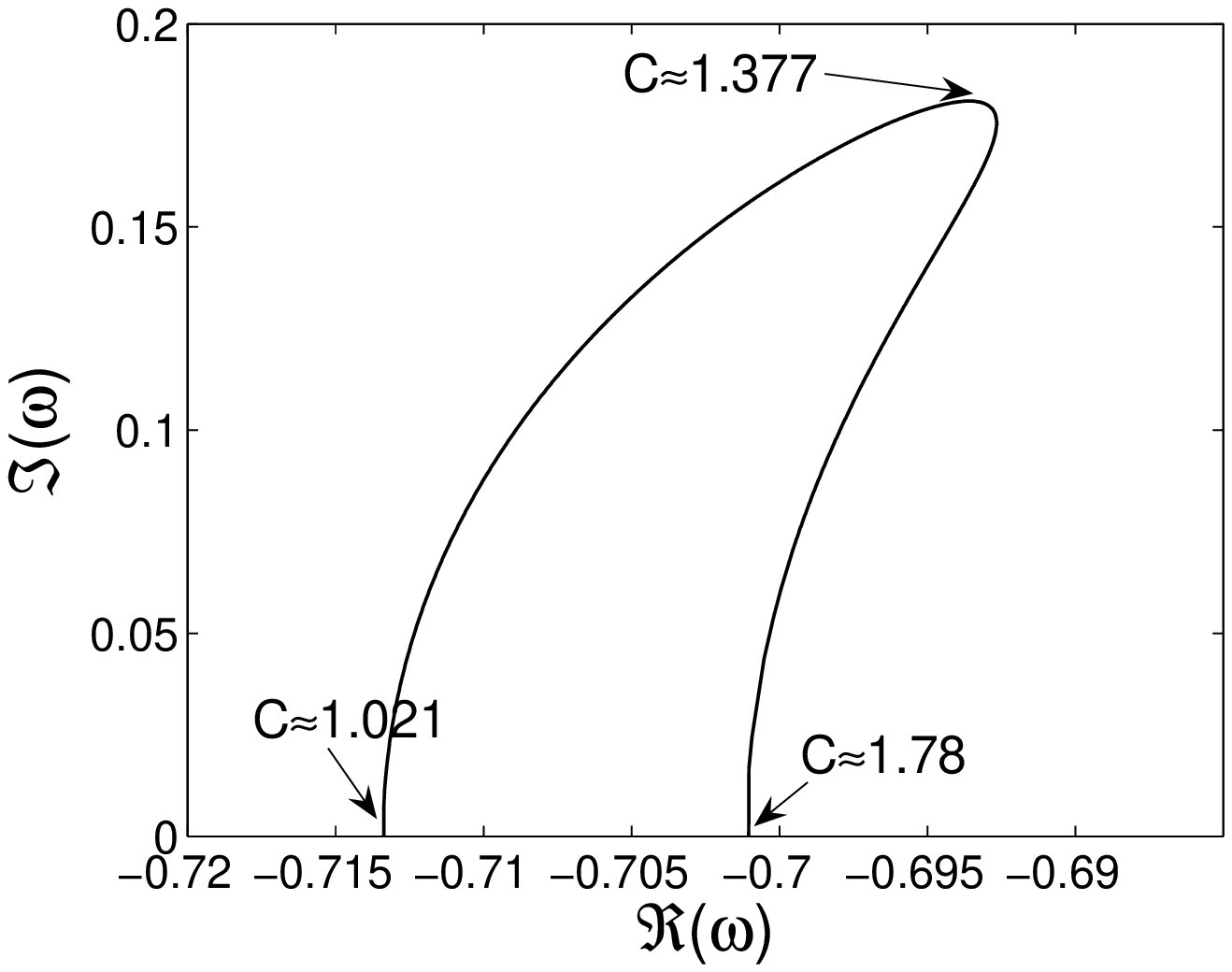}}
\caption{The eigenvalue structure of a single-hump solitary wave 
for $\gamma=0.02$ and $C=1.0$ (top left panel), as well as 
$C=1.4$ (top right panel).
The bottom panel shows the trajectory of one of the unstable eigenvalues 
as $C$ changes.} 
\label{Fig3}
\end{figure}

\begin{figure}[tb]
\centerline{
\includegraphics[height=6cm,angle=0,clip]{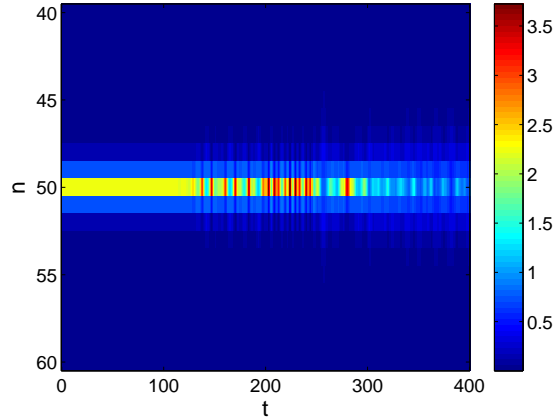}
}
\caption{(Color online) The spatio-temporal evolution of an 
unstable single-hump solitary wave at $\gamma=0.02$ and $C=1.4$.
The contour plot of square modulus $|\phi_n|^2$ is shown.} 
\label{Fig4}
\end{figure}

If $\gamma={\cal O}(C)$, Eq.\ \eqref{wg} becomes
\begin{equation}
\omega = \pm2\sqrt{\Lambda\gamma}\sqrt{C}+{\cal O}(C).
\end{equation}

Next, we have to proceed with calculating the continuous spectrum of the operator ${\cal L}_+(C){\cal L}_-(C)$ \eqref{evp}. 
When $C=0$, all the continuous spectrum of the operator lies at 
$\Omega=\Lambda^2-\gamma^2$ as was mentioned before. When 
$C$ is increased, the eigenvalues spread on the imaginary axis 
creating a phonon band. Using a plane wave expansion 
$\eta_n=ae^{i\kappa n}+be^{-i\kappa n}$ yields the dispersion relation
\[
\Omega=(\Lambda+\gamma+2C-2C\cos\kappa)(\Lambda-\gamma+2C-2C\cos\kappa).
\]
Hence, the continuous band lies between $\Omega_L=\Lambda^2-\gamma^2$ (when $\kappa=0$) and $\Omega_U=\Lambda^2-\gamma^2+8C(\Lambda+2C)$ (when $\kappa=\pi$).

For small $\gamma$, the instability of a one-site discrete breather of \eqref{schr} is caused by the collision of the smallest eigenvalue \eqref{Wg} with an eigenvalue bifurcating from $\Omega_L$. However, here we assume that the bifurcating eigenvalue does not move very fast in the spectral plane such that it can be represented by $\Omega_L$. For large $\gamma$, 
the instability is due to the collision of the smallest eigenvalue and $\Omega_L$. Equating those quantities will give the critical $\gamma$ as a function of the coupling constant $C$, i.e.
\begin{eqnarray}
\gamma^1_{\text{cr}}&=&-\frac25\Lambda-\frac45 C+\frac15\sqrt{9\Lambda^2+16C(\Lambda+C)},
\label{gc1}
\\
\gamma^2_{\text{cr}}&=&-\frac25\Lambda-\frac45 C+\frac15\sqrt{9\Lambda^2+56C\Lambda+96C^2}.
\label{gc2}
\end{eqnarray}
The two approximate $\gamma^i_{\text{cr}}$ above coincide at $C=0$ and $\gamma^*=(\sqrt2-1)/2\approx0.2071$. Notice that at that level the relevant
calculation is analytically exact (i.e., there is no approximation
and the solitary excitation will be unstable for $C=0$ only for
$\gamma=\gamma^*$.


{\it Numerical results.} 
We now proceed to testing our analytical results for the parametrically
driven discrete nonlinear Schr{\"o}dinger system numerically. We start
by examining the validity of our analytical prediction for the 
eigenfrequency corresponding to the phase mode which bifurcates
from $\omega=0$ because of the presence of the parametric drive 
according to the expression (\ref{wg}). Figure \ref{Fig1} shows
this prediction as a function of $C$ for two different values of
$\gamma$. Clearly the prediction is fairly accurate for small $C$
and its range of validity is wider for smaller values of $\gamma$.

We now turn to the examination of the two parameter-plane of
the parametric drive $\gamma$ versus the coupling strength $C$. Figure
\ref{Fig2} provides a full description of the dynamics of the 
parametrically driven DNLS model regarding the intervals of 
stability/instability of the most fundamental, single-hump 
solitary wave solution of the model. The solid lines show the
numerically obtained separatrices between the stable and unstable
parametric regimes of the model, while the dashed and dash-dotted
lines give the analytical prediction for the stability range
as obtained by the conditions of collision of the phase mode
eigenfrequency with the continuous spectrum from Eqs. (\ref{gc1})-(\ref{gc2}).
We observe that the prediction of Eq. (\ref{gc2}) is in very good
agreement with the numerical observations for the occurrence of the
instability point. This is because the collision typically occurs
indeed with the upper band edge of the continuous spectrum (rather
than with an eigenvalue bifurcating from it) and also typically
the collision occurs for small $C$ for which the analytical approximation
of Eq. (\ref{wg}) is a very good approximation. On the other hand,
the slightly less satisfactory agreement with the prediction of 
Eq. (\ref{gc1}) occurs due to the collision with eigenvalues bifurcating
from the lower edge of the continuous spectrum (see also Fig. \ref{Fig3}
below) and also for relatively large $C$'s for which higher order
terms in the expansion of (\ref{wg}) should be expected to contribute.

Figure \ref{Fig3} illustrates the typical instability scenario
for weak parametric drives ($\gamma=0.02$ in this figure). 
As $C$ increases, the eigenvalue
which is associated with the phase mode moves towards the continuous
spectrum (top left panel). 
Eventually for $C \approx 1.021$ it collides with an
eigenvalue pair that has bifurcated from the lower band edge of
the continuous spectrum. Due to the opposite Krein signature of
these eigenvalues (see e.g. the relevant discussion in \cite{kks}),
their collision leads to an oscillatory instability and the bifurcation
of a complex quartet of eigenvalues (top right panel). Eventually,
as is shown in the bottom panel, the eigenfrequencies return to 
the real axis to re-stabilize the configuration for $C>1.78$.

One can also notice from Fig.\ \ref{Fig2} that there is a minimum
$\gamma_m$ below which the soliton is stable all the way to the continuum 
limit. Numerically, $\gamma_m\approx0.0135$. 
When $\gamma$ is less than $\gamma_m$ the eigenvalue (\ref{Wg}) that 
moves towards the continuous spectrum does not collide with the eigenvalue 
bifurcating from the phonon band $\Omega_L$. 
Instead, it
decreases before the collision occurs. This shows that the second 
order correction ${\cal O}(C^2)$ of (\ref{Wg}) dominates the leading
order expression.

We now turn to the examination of the dynamical behavior of the
unstable solutions obtained above. The direct numerical evolution
of an unstable solution of Eq. (\ref{schr}) is shown in Figure \ref{Fig4}.
We have confirmed that this dynamics is typical of the unstable 
parameter range. The figure shows that eventually the solution 
becomes subject to the oscillatory instability that was illustrated
in Fig. \ref{Fig3} and is ultimately destroyed completely. This may
also be expected on the basis of the fact that this is the fundamental
coherent structure solution and for the same parameter set there appears
to be no other stable dynamical state (other than $\phi_n=0$) to which the
initial condition may transform.


{\it Conclusions.} 
In this short communication, we visited the topic of parametrically
driven lattices of the nonlinear Schr{\"o}dinger type. We have shown
that the dynamics of these lattices is considerably different than
those of the regular DNLS equation. This is due to the driving-induced
bifurcation of the phase mode (associated with the gauge invariance
of the NLS equation). Collision of this mode with eigenfrequencies
stemming from the continuous spectrum leads to a wide parametric regime
of instabilities of the fundamental solitary wave in this model. 
Our perturbative analysis captures quite accurately the relevant
eigenvalue (especially for weak couplings) and provides a fair 
estimate of the instability threshold in the parameter-space of
the system. The result of the ensuing oscillatory instability is
the destruction of the fundamental soliton, a feature absent from
the regular DNLS model (where this solution is stable for all parameter
values).

It would be interesting to expand the present considerations to 
other variants of the discrete parametrically driven model such
as its higher-dimensional analogs, the defocusing case, or also
damped variants of these lattice models. Such considerations
are currently under study and will be reported in future publications.


\end{document}